\documentstyle[aps,prl,epsf,multicol,floats]{revtex}

\newcommand{\Imag}{~\mbox{Im}~}
\newcommand{\Real}{~\mbox{Re}~}
\newcommand{\half}{\mbox{\small $\frac{1}{2}$}}
\begin{document}
\draft

\twocolumn[\hsize\textwidth\columnwidth\hsize\csname @twocolumnfalse\endcsname

\title{The Quantum Toda Lattice Revisited}

\author{Rahul Siddharthan\cite{rsidd} and B. Sriram Shastry\cite{bss}}
\address{Department of Physics, Indian Institute of Science, 
          Bangalore 560012, India}

\date{\today}

\maketitle 

\begin{abstract}
In this work we study the quantum Toda lattice, developing 
the asymptotic Bethe ansatz method first used by Sutherland.
Despite its known limitations we find, on comparing with 
Gutzwiller's exact method, that it works well in this 
particular problem and in fact becomes exact as $\hbar$ 
grows large. We calculate ground state and excitation 
energies for finite sized lattices, identify excitations 
as phonons and solitons on the basis of their quantum
numbers, and find their dispersions. These
are similar to the classical dispersions for small $\hbar$,
 and remain similar all the way up to $\hbar=1$, but then 
deviate substantially as we go further into the quantum 
regime. On comparing the sound velocities for various 
$\hbar$ obtained thus with that predicted by conformal 
theory we conclude that the Bethe ansatz gives the energies
per particle accurate to O($1/N^2$). On that assumption 
we can find correlation functions. Thus the Bethe ansatz 
method can be used to yield much more than the 
thermodynamic properties which previous authors have 
calculated.
\end{abstract}


\vskip2pc]

\flushbottom

\bibliographystyle{unsrt}

\section{Introduction}

The Toda lattice \cite{todabook}, introduced by M. Toda in 1967
\cite{todapaper}, is a chain of particles which interact with 
nearest neighbours with an exponential potential.
The quantum mechanical Hamiltonian for a periodic Toda system of
length $N$ (i.e. $n+N \equiv n$) is
\begin{equation}
\label{toda_ham}
  H = - \sum_{n=1}^N \frac{\partial^2}{\partial {u_n}^2}
     + \eta \sum_{n=1}^N e^{-(u_{n+1} - u_n)}
\end{equation}
where the $u_n$ are displacements from equilibrium sites. We have
chosen appropriate units to remove $\hbar$, $m$ (the mass of the
particle) and the length scale of the potential. The infinite system
also has a linear term in the potential (to cancel the one in the
exponential), but with periodic boundary conditions this vanishes.
$\eta$ is a measure of the anharmonicity and also of the scale of the
quantum effects.  The larger $\eta$ is, the more ``classical'' the
system  and the more harmonic the low-energy excitations.
In the classical limit the parameter
$\eta$ can be scaled out but in the quantum case this can only be
done by introducing an $\hbar \neq 1$ in the above equation. We shall
occasionally write
\begin{equation}
 \label{hbareta}
 \hbar = \sqrt{\frac{2}{\eta}}
\end{equation}
so that the Hamiltonian can be rescaled and rewritten as
\begin{equation}
 \label{todaham_hbar}
  \frac{\hbar^2}{2} H = - \frac{\hbar^2}{2} \sum_{n=1}^N 
  \frac{\partial^2}{\partial {u_n}^2} + \sum_{n=1}^N e^{-(u_{n+1}-u_n)}.
\end{equation}

\begin{figure}
  \begin{center}
  \leavevmode
  \epsfxsize=6.5cm
  \epsfbox{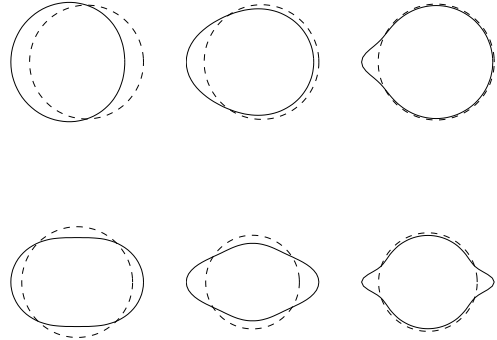}
  \end{center}
  \caption{How a classical cnoidal wave continuously goes from nearly
           harmonic to solitonic with increasing amplitude. Above, the first
           ``normal mode'' which goes into a one-soliton state. Below, the
           second mode which goes into a two-soliton state. The same thing
           happens in our quantum description when we put more and more
           phonons into a particular phonon mode. (Not to scale)
   \label{cnoid}}
\end{figure}

The Toda lattice is interesting, classically and quantum
mechanically, because it is the one example of a nonlinear lattice
which can be solved exactly. Elementary excitations are cnoidal
waves, which are periodic waves analogous to the normal modes of a
harmonic lattice, and solitons, which are travelling pulse-like
solutions which retain their shape even after interaction with other
excitations. The periodic system does not support solitons of the
infinite-chain type,
since these involve a net compression, but a cnoidal wave with large
amplitude behaves very much like a soliton (Fig.\ \ref{cnoid}).

The classical periodic system was studied by Kac and van Moerbeke,
and Date and Tanaka \cite{KvM,date}. It is completely
solved \cite{todabook}:
given any initial condition of the system its future time evolution
can be written down exactly.
In quantum mechanics, there have been several treatments based on
various approximations and assumptions.  On the one hand, Gutzwiller 
\cite{gutz}
has given an exact treatment of the 3 and 4 particle lattices, and
his quantization algorithm is capable of generalization to larger $N$
as well. His results were rederived in the $r$-matrix formalism
by Sklyanin \cite{skly} and by Pasquier and Gaudin \cite{pasq-gaud}. 
The method is plausible and makes a transparent connection with the classical
formulation of the problem. On the other hand calculating with this
algorithm is a formidable task. The method is summarized in
\S \ref{sec:Gutzwiller}.
On the other hand,
some authors have used the (asymptotic) Bethe ansatz to treat the problem.
Sutherland \cite{suth}
originally recovered the classical results (high $\eta$) in the
thermodynamic limit ($N \rightarrow \infty$). Later authors \cite{mert}
have remained in this thermodynamic limit, but have looked at
arbitrary $\eta$, and have calculated various thermodynamic 
functions. 

In this paper, we use the Bethe ansatz to look at the case of finite
$N$, which in some ways is more illuminating when one tries to
classify excitations as phonons or solitons. 
\S \ref{sec:scaling} obtains the Toda lattice as a limiting case of the
$1/\sinh^2$ model, an idea due to Sutherland, and \S \ref{sec:Bethe_solution}
sets up the Bethe Ansatz equations for the latter model and performs
the same limit to obtain equations describing the Toda lattice.
Though the asymptotic Bethe ansatz is in general inaccurate for finite $N$,
we find in this model that it is much better than it has been given credit
for, and in particular becomes exact not only for $N \rightarrow \infty$
but also for $\eta \rightarrow 0$ with finite $N$. In \S 
\ref{sec:Gutzwiller} we demonstrate this by setting up exact equations
using Gutzwiller's method and seeing what approximations are involved in
going from these to the Bethe equations. The claim \cite{fowlfr} 
that the Bethe ansatz misses a fixed fraction of
states does not stand scrutiny. One need only glance at the harmonic limit
(\S \ref{sec:harmonic}) where every one of the states is accounted 
for accurately. 

Having done this, we examine the opposite, highly quantum
limit in \S \ref{sec:anharmonic}, which makes clearer how the low lying
phonon-like modes go over to soliton-like states as their occupation number
is increased. \S \ref{sec:dispersion} calculates dispersion relations
for phonons and solitons, and compares the classical and quantum results.
We find that the results are essentially the same, apart from the quantization
of energy levels, for $\eta>2$ ($\hbar<1$) roughly; but as one decreases $\eta$
further the quantum results deviate more and more
from the classical, though they remain qualitatively similar down to $\eta
\rightarrow 0$. In this regime we get phonon-like excitations whose energies
cannot be derived from harmonic approximations (why we think of them as
phonons is discussed in \S \ref{sec:dispersion}), and soliton-like excitations
which can be thought of as authentic examples of the much discussed ``quantum
soliton''.

\S \ref{sec:EvsN} considers finite size effects and makes contact with
conformal theory to find correlation functions. We offer evidence that
the asymptotic Bethe ansatz, in this problem, gives the energy per particle
accurately to order $1/N^2$, though on general grounds it is guaranteed 
only to give results accurate to order 1. Finally, we consider in
\S \ref{sec:classical} how all this relates to the classical lattice,
and the appendix gives, for completeness, a brief discussion of the 
other conserved quantities
(H\'{e}non's integrals) and why they are conserved in the classical and
quantum cases.

\section{Scaling the $1/\sinh^2$ model to the Toda model} 
\label{sec:scaling}

Sutherland was the first to treat the quantum Toda lattice, as a
limiting case of the $1/\sinh^2$ model, by pioneering the use
of  the asymptotic Bethe ansatz.
 He contented himself with recovering the classical results, and 
showed that the  classical solitons are recovered by taking
the classical limit of particle like excitations of the 
Bethe equations. He did not explore regimes other than the classical,
thermodynamic limit.  Later
authors like Mertens \cite{mert} have directly treated the Toda lattice by
Bethe's ansatz, using the phase shifts obtained from the Toda
potential; but the validity of the Bethe ansatz (which involves
summing over phase shifts a given particle suffers in collisions with
all other particles) is unclear in a model where only
nearest-neighbour interactions appear. We therefore use Sutherland's
approach and scale the $1/\sinh^2$ model. Our scaling procedure
is somewhat more explicit and displays the fact that the
limiting process leaves us with a one parameter model, the Toda
lattice with a general coupling constant $\eta$ ( see below), 
from which the classical,
the harmonic, and the extreme anharmonic limits follow.

Our starting point is the Hamiltonian
\begin{equation}
\label{sinh_ham}
  H_S = - \sum_{n=1}^N \frac{\partial^2}{\partial {x_n}^2}
     + g \sum_{{m,n=1} \atop {m<n}}^N \frac{1}{\sinh^2[(x_m - x_n)/2a]}.
\end{equation}
Here $a$ is a length scale giving the range of the potential,
$g$ is a coupling constant and the particles are on a ring of
length $L$ (so that the density $d=N/L$).  In the dilute limit
when the particles are far apart, the $\sinh^2$ becomes an
exponential; we achieve this limit by making the substitution
\begin{equation}  
  x_n = n/d + u_n a
\end{equation}
where $u_n$ are displacements from lattice sites spaced $1/d$ apart.
We let $ad$ go to zero, and assume that the $u$'s are bounded (that
is, the wavefunction vanishes as $u \rightarrow \infty$).
Then we have for  $m<n$ and $da \ll 1$,
\begin{eqnarray}
  \sinh^2 \left(\frac{x_m-x_n}{2a} \right)& =&\sinh^2 \left( \frac{m-n}{2ad} +
                      \frac{u_m-u_n}{2} \right) \nonumber \\
&=& (1/4) \exp \left( \frac{n-m}{ad} + u_n-u_m \right) \nonumber
\end{eqnarray}
 and the potential in the Hamiltonian becomes
\begin{equation}
  4g \sum_{m<n} \exp \left[ -\frac{n-m}{ad} - (u_n-u_m) \right].
\end{equation}
So on putting
\begin{equation}
  \label{g_to_eta}
  g = \frac{\eta}{4a^2} e^{1/ad}
\end{equation}
and then allowing $ad$ to go to zero, all terms in the interaction except
the nearest-neighbour terms (i.e. $n=m+1$) are killed, and we finally arrive
at the Hamiltonian
\[
a^2 H_S  =  - \sum_{n=1}^N \frac{\partial^2}{\partial {u_n}^2}
     + \eta \sum_{n=1}^N e^{-(u_{n+1} - u_n)}
\]
which is the Toda Hamiltonian, eq.\ (\ref{toda_ham}).

\section{Solution by the Asymptotic Bethe Ansatz}
\label{sec:Bethe_solution}

The $1/\sinh^2$ system, being integrable, is characterised by $N$
commuting integrals of motion.
If we suppose that the particles are moved far away from one another,
they don't interact except during short-range collisions, and for
the rest of the time they have well-defined momenta which
can be taken to be the conserved quantities. During two body collisions
the most that can happen is an exchange of momenta, and
one can show that $n$-particle collisions can be completely described in
terms of successive 2-particle collisions and their phase shifts, so
that the momenta are reordered but not changed.
The Bethe Ansatz is a sum of plane-wave product states, characterized by
a set of single-particle momenta \{$p_n$\} and an amplitude for
each plane wave state which features a different permutation of these
momenta. All that is required for calculation is the two-body
phase shift $\theta(p-p')$ for two particles with momenta $p$ and
$p'$ \cite{suth}; then equations for the $p_n$ can be written down
and solved. These equations are
\begin{equation}
  \label{bethe_eqn}
   p_n = \frac{2 \pi}{L} I_n + \frac{1}{L} \sum_{{m=1} \atop {m \neq n}}^N
        \theta (p_n-p_m).
\end{equation}
Here $I_n$ are integers
for odd $N$, or half-odd-integers for even $N$, no two of which are
equal.  The energy (eigenvalue of $H_S$) is then given by $\sum
{p_n}^2$, so that the energy of the corresponding Toda problem would
be $\sum a^2{p_n}^2$.

This solution is derived in the limit when the particles are far
apart, weakly interacting and in approximately plane wave states, so
na{\"\i}vely one would not expect the results to hold at higher densities.
It is known, however, that this ``asymptotic Bethe ansatz'' holds at
all densities in the limit $N \rightarrow \infty$ (the thermodynamic
limit) provided the virial expansion has no singularities as a
function of $d$ \cite{panch}. For this particular problem, it turns out that
the solution is also exact for arbitrary $N$ in the limit $\eta
\rightarrow 0$. Otherwise, though not exact, it is often a very good
approximation.

The total momentum is 
\begin{equation}
  \label{ktot}
  p_{\rm tot} = \frac{2 \pi d}{N} \sum_{n=1}^N I_n .
\end{equation}
Owing to the Galilean Invariance of the $1/\sinh^2$ model, a state with
zero total momentum can always be boosted to have a total momentum
$p_{\rm tot}$ by adding an appropriate integer to all the quantum numbers, 
at a net energy cost $p_{\rm tot}^2/N$ independent of the coupling constant.
We can take the expression for the two-body phase shift in the
$1/\sinh^2$ system from Sutherland:
\begin{equation}
  \label{sutherland_phshift}
  \theta(p) = 2 [\arg \Gamma (1+S+ipa) - \arg \Gamma (1+ipa)]
\end{equation}
where $S(S+1) = 2ga^2 = \frac{\eta}{2} e^{1/ad}$.
Since we are taking the dilute limit, $S \rightarrow \infty$ for any
value of $\eta$, and we can write
\begin{equation}
  \label{S}
  S = \sqrt{\frac{\eta}{2}} e^{1/2ad}.
\end{equation}
In the limit $S \rightarrow \infty$, the phase shift
(\ref{sutherland_phshift}) becomes 
\begin{equation}
  \label{our_phshift1}
  \theta(p) = 2 pa \log S - 2 \Imag \log \Gamma(1+ipa)
\end{equation}
[we can show this by using Stirling's expansion for large $S$ in
the first gamma function in (\ref{sutherland_phshift})].
We substitute for $S$ from (\ref{S}), put the resulting phase
shift into the Bethe equations (\ref{bethe_eqn}), noting that
$\sum_m (p_n-p_m) = N p_n-p_{\rm tot}$
where $p_{\rm tot}$ is given by (\ref{ktot}), and rearrange [the
$p_n$ on the left of (\ref{bethe_eqn}) cancels with a term from
the phase shift, leaving only O($1/L$) and smaller terms]. Defining
dimensionless ``momenta'' by $k_n=p_n a$, dividing
out the common $d$ and taking $ad \rightarrow 0$
we end up with the equations to be solved:
\begin{eqnarray}
   \alpha k_n & = &
   -\frac{\pi}{N} \left( I_n - \frac{\Sigma I_n}{N} \right) \nonumber \\
 & & + \frac{1}{N} \sum_{{m=1} \atop {m \neq n}}^N 
                 \Imag \log \Gamma [1+i(k_n-k_m)]   \label{bethe_eqns_final}
\end{eqnarray}
where for convenience we have written
\begin{equation}
  \label{alpha}
  \alpha = \frac{1}{2} \log \left( \frac{\eta}{2} \right) (= - \log \hbar).
\end{equation}
Note that the total momentum of the system, 
$k_{\rm tot} = (a d) (2 \pi /N) \sum_j I_j$, goes to zero as $ad$ goes
to zero, so that we are working in a zero momentum frame. This is a consequence
of the length of the underlying $1/\sinh^2$ model going to infinity (on
the scale of the range of the potential), the 
momentum being inversely proportional to the system length. However, our
simultaneous scaling up of the interaction by an exponential factor 
(\ref{g_to_eta}) ensures that the individual particle momenta remain finite.
Thus we have gone from a $1/\sinh^2$ ``gas'' with particles described by
actual position coordinates, to a lattice with particle positions described
as displacements from lattice sites, and no net momentum, which is what we
wanted. The energy of this Toda problem is $\sum k^2_n$.
 Since the problem continues to be Galilean invariant, a
finite momentum $k_{\rm tot}$ can always be introduced into 
the above equations by adding $\alpha k_{\rm tot}/N$ to the right hand side, 
at a total energy cost of $k^2_{\rm tot}/N$. This $k_{\rm tot}$ need not be 
quantized, since as the length of the underlying $1/\sinh^2$ model expands 
the quanta of momentum become infinitesimal.

The $I_n$ in (\ref{bethe_eqn}) and (\ref{bethe_eqns_final}) are the
quantum numbers of the system, and uniquely
specify the state of the system.
The momenta $k_n$ are ordered in the same way
as $I_n$ [despite the apparently opposite sign for $\eta>2$ in
(\ref{bethe_eqns_final})] and we assume that the order is ascending in
$n$. In the ground state the $I_n$ are successive
integers (or half integers), generally taken to be centred about zero
[though it does not matter here, since one subtracts their
average value in Eq.\ (\ref{bethe_eqns_final})] and in the excited
states one or more of them are increased by various integer values,
always making sure no two of them have the same value.

Although we took the dilute limit in arriving at these equations, the
Toda Hamiltonian (\ref{toda_ham}) which they describe contains no
reference to the lattice constant, and therefore they are valid at
all densities, or at least at all densities sufficiently low that the
particles do not cross each other. (The wavefunction will give the
typical ``spread'' in $u_n$ and we must assume, for physical reasons,
that the inter-particle separation is much larger than this).
Mertens' treatment \cite{mert}, if followed through, 
gives the same equations as the
above but with an extra term on the right hand side equal to
$(2\pi ad/N) \sum I_n$ (which is the above $k_{\rm tot}$: 
he does not consider a limiting case of the
$1/\sinh^2$ model and does not take $d \rightarrow 0$). This term has
no significance and, in particular, must not be
confused with the phonon or soliton momenta (\S \ref{sec:dispersion}).
In fact, it may be subtracted out, since it is independent of $n$, to
recover our equations. We prefer this, the rest frame, because it is
the frame in which one normally discusses phonons and also because it is
convenient in making contact with Gutzwiller's work.

Since one can add a constant quantity to the $I_n$ without effect on
the equations, they contain some redundancy---$N-1$ quantum numbers
are enough to characterize the system. We could define new
quantum numbers by
\begin{equation}
 \label{nunumbers}
  \nu_n = I_{N-n+1} - I_{N-n} - 1, ~~~n=1, 2, \ldots, N-1
\end{equation}
so that the $\nu_n$ may take any integer value from 0 upwards.
(These are the number of ``holes'' between successive integers $I_n$,
starting from the right.) These are, in the harmonic limit, the phonon
occupation numbers (\S \ref{sec:harmonic}).

Equations (\ref{bethe_eqns_final}) can be solved numerically, for
instance by the Newton-Raphson method, for moderate values of $N$
without much difficulty if one has a good starting guess. If not,
the numerical methods tend to converge to spurious
solutions where the ordering of the $k$'s is not the same as that of
the $I$'s.

Alternatively, one could pass to the thermodynamic limit and write
down integral equations from which various thermodynamic quantities
could be calculated, as in Yang and Yang's treatment of the
$\delta$-function Bose gas. This has been done by Mertens, and by
Hader and Mertens \cite{mert}. We define $(N/2\pi ) \xi (k) dk$
as the number of $k$'s between $k$ and $k+dk$. Then
(\ref{bethe_eqns_final}) yields the
integral equation for the density of the $k$'s in the ground state
which is, in agreement with Mertens,
\begin{equation}
\xi(k) = - 2\alpha + \frac{1}{\pi} \int_{-B}^{B}
             \xi(k') \Real \psi(1+i(k-k')) dk'.
\end{equation}
($\psi$ is the digamma function). For reasons given in the next section,
Matsuyama \cite{matsu96} gets the same equation for the distribution of the
zeros of Hill's determinant in the Gutzwiller method
(but without the inhomogeneous part since he takes $\hbar=1$, or $\eta=2$.)

\section{Comparison with Gutzwiller's formulation}
\label{sec:Gutzwiller}

The Bethe equations for the Toda lattice can also be derived from
Gutzwiller's solution of the problem, if some approximations are
made. This helps clarify what the $k$'s mean in the
non-dilute limit---in particular, their correspondence with the
classical variables, and also tells us when our approximations are
valid. We briefly describe Gutzwiller's method and the resulting
quantization conditions.

Gutzwiller, following the classical ideas of Kac and van Moerbeke
\cite{KvM}, tries to write the wavefunction of the $N$ body lattice
as a series involving the wavefunctions of the $N-1$ body open
lattice obtained by removing one particle.
Suppose these $N-1$ body wavefunctions are $\Psi_{\kappa_1 \kappa_2
\ldots \kappa_{N-1}}$; the indices $\kappa$ correspond to the
classical variables $\mu_i$ (the eigenvalues of the truncated $N-1$
dimensional Lax matrix). For the open chain they are purely imaginary
but when using them as a basis in the closed chain Gutzwiller shows
that one must extend them to have a real integer part; in other
words, $\kappa_i = i \rho_i + k_i$, where $k_i$ is an integer.
One aims to find the spectrum of the $\rho$'s. It turns out that if
one writes the wavefunction as $\Psi = \sum C_{\kappa_1 \kappa_2
\ldots \kappa_{N-1}} \Psi_{\kappa_1 \kappa_2 \ldots \kappa_{N-1}}$,
where the sum is over the integers $k_i$, one can get a solution of
the form $C_{\kappa_1 \kappa_2 \ldots \kappa_{N-1}} =
(\kappa_1 - \kappa_2)(\kappa_2 - \kappa_3) \ldots r_{\kappa_1}
s_{\kappa_2} t_{\kappa_3} \ldots$ provided the coefficients
$r$, $s$, $\ldots$ satisfy identical recursion relations
\begin{equation}
i^N r_{\kappa+1} + i^{-N} r_{\kappa-1} = D(\kappa) r_\kappa
\end{equation}
where $D(\kappa)$ is basically the characteristic polynomial of the
Lax matrix (see the appendix):
\begin{eqnarray}
 D(\kappa) &=& \hbar^N \kappa^N +E \hbar^{N-2} \kappa^{N-2}
                 +i A_3 \hbar^{N-3} \kappa^{N-3} + \nonumber \\
           & &   \ldots + (-i)^{N-1} A_{N-1} \hbar \kappa + (-i)^N A_N,
\end{eqnarray}
and $\hbar$ is defined in (\ref{hbareta}).
Suppose that its zeroes are $i\epsilon_1, i\epsilon_2, \ldots,
i\epsilon_N$; then $D(\kappa)$ can also be written as
\begin{equation}
D(\kappa) = \prod_n (\hbar \kappa - i\epsilon_n).
\end{equation}

The same recursion relations are derived by Sklyanin, and by Pasquier
and Gaudin, from different points of view \cite{skly,pasq-gaud}.
They have two independent solutions, differing in their behaviour at
$+\infty$ and $-\infty$. Gutzwiller sets
\begin{eqnarray}
\label{rkappa12}
r_\kappa^{(1)} = \frac{(-1)^\kappa r'}{\hbar^{N\kappa} \prod_i 
  \Gamma(1+\kappa-i\epsilon_i)} r'_\kappa, \\
r_\kappa^{(2)} = \frac{(-1)^\kappa r''}{\hbar^{-N\kappa} \prod_i 
  \Gamma(1-\kappa+i\epsilon_i)} r''_\kappa, \nonumber
\end{eqnarray}
where $r'$ and $r''$ are coefficients to be matched later when
``joining'' the two solutions, and $r'_{\kappa}$ and $r''_{\kappa}$
are two new variables which (it turns out) are complex conjugate.
They have solutions
\begin{eqnarray}
 \label{r_kappa}
 r'_\kappa = \left|
  \begin{array}{cccc}
  1 \; & \frac{\pm 1}{D(\kappa+1)} &  & 0 \\ [0.5em]
  \frac{1}{D(\kappa+2)} \; & 1 \; & \frac{\pm 1}{D(\kappa+2)} & \\ [0.5em]
  & \frac{1}{D(\kappa+3)} \; & 1 \; & \\ [0.5em]
  0 & & & \ddots
  \end{array}
 \right|, \\ [1em]
r''_\kappa = \left|
  \begin{array}{cccc}
  \ddots \; & & & 0 \\ [0.5em]
  & 1 \; & \frac{\pm 1}{D(\kappa-3)} & \\ [0.5em]
  & \frac{1}{D(\kappa-2)} \; & 1 \; & \frac{\pm 1}{D(\kappa-2)} \\ [0.5em]
  0 & & \frac{1}{D(\kappa-1)} \; & 1
  \end{array}
 \right|.
\end{eqnarray}
The former approaches a constant as $k \rightarrow +\infty$, and the
latter approaches a constant as $k \rightarrow -\infty$.

If one tries to join these solutions, one gets the consistency condition
\begin{equation}
 \label{hill}
\Delta(\kappa) = \left|
   \begin{array}{ccccc}
    \ddots & \frac{\pm 1}{D(\kappa-2} \; & & & 0 \\ [0.5em]
   &  1 \; & \frac{\pm 1}{D(\kappa-1)} & &\\ [0.5em]
    & \frac{1}{D(\kappa)} \; & 1 \; & \frac{\pm 1}{D(\kappa)} &\\ [0.5em]
   & & \frac{1}{D(\kappa+1)} \; & 1 \; &\\ [0.5em]
   & & & \frac{1}{D(\kappa+2)}  & \\ [0.5em]
    0 & & & & \ddots
   \end{array}
 \right| = 0.
\end{equation}
Here the $+$ signs are for even $N$, the $-$ signs for odd $N$.
This determinant has $N$ purely imaginary zeroes, which we call
$i\rho_1, i\rho_2, \ldots, i\rho_N$ (in ascending order). (If $N$ is
odd and all odd integrals $A_3, A_5, \ldots$ vanish---this happens,
for instance, in the ground state---then there are only $N-1$ zeroes
but in that case $\kappa=0$ also satisfies the quantization
conditions below, so we include it among the $\rho$'s.) It is clear
that in addition to these, $i\rho_n+l$, where $l$ is an arbitrary
integer, are also zeroes of the determinant.

The determinant is part of what we need to find the spectrum of
$\rho$, but it is not enough since we don't know the constants of
motion in $D(\kappa)$. We need more quantization conditions; to
supply these
Gutzwiller defines an angle $\phi = (1/2)\arg(r'/r'') =
\arg(r')$ since $r'$ and $r''$ are complex conjugate. If one
normalizes the solutions by $r_{i\rho}^{(1)} = r_{i\rho}^{(2)} =
1$, one finds
\begin{equation}
\phi = \arg(r') = \Imag \log \left( \frac{\hbar^{iN\rho} \prod_m
          \Gamma (1+i[\rho-\epsilon_m])}{{r'_{i\rho}}} \right).
\end{equation}
Then $\phi$ is a monotonically increasing function of $\rho$. Abbreviating
$\phi(\rho_n)$ as $\phi_n$, Gutzwiller's quantization condition reads
\begin{equation}
 \label{phaseequal}
\phi_1 = \phi_2 = \ldots = \phi_N ~~~~ (\mbox{modulo}~~ \pi).
\end{equation}
In addition he assumes that
\begin{equation}
 \label{phasesum}
\phi_1 + \phi_2 + \ldots + \phi_N = 0.
\end{equation}
If both of these conditions are satisfied, the allowed values of $\phi_n$
are very limited; they can only be of the form $I_n\pi + m\pi/N$,
where $m$ is the same integer for all $n$ and $I_n$
is an arbitrary integer, different for different $n$. But $\phi_n$ is
an increasing function of $\rho_n$, hence if the $\rho_n$ are
ordered, we must have the $I_n$ also in increasing order. Then, from
(\ref{phasesum}), we get
\begin{equation}
 \sum_n \pi I_n + m \pi = 0
\end{equation}
which yields $m=-\sum I_n$. So we have, finally, expressions for
Gutzwiller's phase angles:
\begin{eqnarray}
\phi_n & = & \pi \left(I_n - \frac{\sum I_m}{N}\right) \nonumber \\
        & = & \arg r' = \arg \left( 
    \frac{\hbar^{iN\rho_n} \prod_m \Gamma (1+i[\rho_n-\epsilon_m])}{
       r'_{i\rho_n}} \right) \nonumber \\
& = & -\alpha N \rho_n + \sum_m \Imag \log \Gamma
             (1+i[\rho_n-\epsilon_m]) \nonumber \\
             & & - \Imag \log r'_{i\rho_n},
\end{eqnarray}
or
\begin{eqnarray}
\alpha \rho_n & = & -\frac{\pi}{N}\left( I_n - \frac{\sum I_m}{N} \right)
 \nonumber \\
 & &  + \frac{1}{N} \sum_m \Imag \log \Gamma(1+i[\rho_n-\epsilon_m]) 
   \nonumber \\
   & &      - \frac{\Imag \log r'_{i\rho_n}}{N}. \label{gutzfinal}
\end{eqnarray}
($\alpha = \half \log(\eta/2) = -\log \hbar$.)
These, then, are the exact Gutzwiller equations which can be combined
with (\ref{hill}) to calculate the $\rho_n$ and $\epsilon_n$; once
the latter are known, all the conserved quantities can be found.
The $I_n$ in this equation are the quantum numbers of the system, and
are the same as the $I_n$ in the earlier, very similar Bethe
ansatz equations (\ref{bethe_eqns_final})---
to which these equations in fact reduce
provided (1) the last term can be ignored and (2) $\rho_n$ is very
close to $\epsilon_n$ for all $n$. These things can happen under
two circumstances.

There is an argument in \cite{matsu96} showing that the $\rho_n$ should
approach $\epsilon_n$ as $N \rightarrow \infty$ (and one knows on
general grounds that the asymptotic Bethe ansatz is correct in this
limit). This also
happens as $\hbar \rightarrow \infty$, for finite $N$. 
We can understand the latter fact intuitively as follows: As $\hbar
\rightarrow \infty$, the polynomials $D(\kappa)$ tend to
infinity. (This is not obvious---for example, they don't vanish
as $\hbar \rightarrow 0$ ---but it will be demonstrated in \S
\ref{sec:anharmonic}). 
Then they will be small only in a
small region close to their zeroes, so the matrix of which
(\ref{hill}) is the determinant tends to the unit matrix except when
$\rho_n$ lie in some small regions surrounding $\epsilon_n$. Thus the
determinant can only vanish when the $\rho$'s approach the
$\epsilon$'s; otherwise it is close to unity. For the same reason,
(\ref{r_kappa}) tends to unity (its zeroes will be close to $i
\rho_n + l = 0$, $l \ge 1$, and for $\kappa=i\epsilon_n$ all the
$D$'s in the denominators will be very large). Then the last term in 
(\ref{gutzfinal}) will vanish, and all the $\rho$'s can be
substituted with $\epsilon$'s, and we recover exactly the Bethe
ansatz equations.

Thus the Bethe ansatz is actually
more accurate in the quantum limit than in the classical limit.
Indeed, even for $\hbar=1$ and $N$ = 4--6, the agreement with Matsuyama's
exact diagonalization results \cite{matsu92} is excellent (one gets exactly his
answers, to his reported accuracy) ---and this looks neither like a
thermodynamic limit nor like an extreme quantum limit.

One might imagine that the Bethe ansatz equations could be improved
by subtracting the term $\frac{1}{N} \Imag \log r'_{ik_n}$, but it
turns out that this term is always small compared to the others and
does not greatly improve the results, while it is computationally
expensive to include; therefore we ignore it in all cases.

Finally, we observe that equations (\ref{gutzfinal}) do not remain
the same if the $\rho$'s and $\epsilon$'s are increased by a constant
quantity, because of the last term which does not appear in the Bethe
ansatz equations. We cannot therefore transform these easily to a
non-zero-momentum frame.

\section{The harmonic limit (high $\eta$)}
\label{sec:harmonic}

For large $\eta$ (the classical limit) the lattice is
harmonic, at least for sufficiently
small quantum numbers. The larger $\eta$ is, the larger the
energies and the quantum numbers required for anharmonicity to show up.
Treating this case makes clear the mapping between the phononic
quantum numbers and the $I_n$.

First, the exact solution. There are $N-1$ normal modes in the
system, characterized by ``phonon momenta'' or wavenumbers
$q_n = 2 \pi n /N$, where $n$ = 1 \ldots $N-1$. In our
notation the coefficient of the $u^2$ terms in (\ref{toda_ham}) is
$\eta /2$. Then the frequency $\omega_n$ of the $n$th
mode is 
\begin{equation}
  \omega_n = 2 \sqrt{2 \eta} \sin \left( \frac{q_n}{2} \right).
\end{equation}

An arbitrary state of the system is then characterised by a set of
nonnegative integers \{ $\nu_n$ \} (phonon occupation numbers). The
energy of such a state is
\begin{eqnarray}
 \label{e-harmonic}
  E & = & N \eta + \sum_{n = 1}^{N-1} (
           \nu_n+ \half ) \omega_n \nonumber \\ 
     & = & N \eta + 2 \sqrt{2\eta} ~ \sum_{n=1}^{N-1}
     (\nu_n+ \half) \sin \left( \frac{\pi n}{N} \right) .
\end{eqnarray}
The first term arises from the constant term in the Taylor expansion
of the exponential potential. For the ground state, we set $\nu_n=0$
and find
\begin{equation}
E= N \eta + \sqrt{2 \eta} \; \cot(\frac{\pi}{2 N}) \label{harmonicexact}
\end{equation}
which for large $N$ has an expansion
\begin{equation}
E= N \eta + \sqrt{2 \eta} \; [ \; \; \frac{2}{\pi} N -
 \frac{\pi}{6N}  +O(1/N^3)]
.\label{harmoniclargen}
\end{equation}
 
Now we compare this expression with the result of solving Eqs
(\ref{bethe_eqns_final}) numerically for 10 particles, and for
various ``large'' $\eta$ : 10, 100, and higher. The ground
state is when the $I_n$ are contiguous, with no ``holes''; the energy
calculated from (\ref{e-harmonic}) is in good agreement with the
value obtained from the Bethe ansatz. Numerical
calculations show that
the $\nu_n$ which describe a phononic state are exactly the numbers
defined in (\ref{nunumbers}). In other words, the number of phonons
in a mode $n$ is given by the number of holes between $I_{N-n}$ and
$I_{N-n+1}$. 

This prescription accounts for all the states of the harmonic
lattice, and the quantitative agreement is very close for low phonon
numbers (the higher $\eta$ is, the higher the allowed phonon numbers
before anharmonic effects start showing up). Fig.\ \ref{phonons} gives the
dispersion curve for single phonons; only for low $\eta$ does it differ from
the harmonic-lattice curve. Calculations show that the energies of phonons
are additive (provided there are not too many of them) so
multi-phonon states are also accurately described.

\section{The strongly-quantum anharmonic limit ($\eta \rightarrow 0$)}
\label{sec:anharmonic}

In the large-$\eta$ case, increasing occupation numbers will bring
out anharmonic corrections in the energy, and modes with very high
occupation numbers will resemble solitons. In \S
\ref{sec:dispersion} we demonstrate this with calculations, but
if $\eta$ is not large anharmonicity shows up even in low-lying modes.

Having looked at the harmonic limit in the last section, we now look at the 
opposite limit of the lattice, $\eta \rightarrow 0$; in this case it
turns out that the phase shift simplifies greatly, and we can in fact
solve equations (\ref{bethe_eqns_final}) for $k_n$---an uncommon
phenomenon in Bethe ansatz calculations. 

Equation (\ref{our_phshift1}) for the phase shift is
\[
  \theta(k) = 2 k \log S - 2 \Imag \log \Gamma (1+ik),
\]
and as $\eta \rightarrow 0$, $k$ also becomes small. In this limit
the term involving the gamma function becomes $2 \gamma k$, where
$\gamma = 0.577215...$ is Euler's constant. A quick way to derive
this result is to assume $S$ is a large integer in
(\ref{sutherland_phshift}) and to expand the first gamma function as
a product $(S+ik)(S-1+ik)\ldots (1+ik) \Gamma (1+ik)$,
and if $k \ll 1$,
the argument of this is $k/S + k/(S-1) + \ldots + k$ + a piece which
cancels the second term in (\ref{sutherland_phshift}). As $S
\rightarrow \infty$, using the definition $
\gamma = \lim_{n \rightarrow \infty} 1 + (1/2) + (1/3) + 
          \ldots + (1/n) - \log n$,
the phase shift becomes
\begin{equation}
  \label{small_eta_phshift}
  \theta(k) = 2 k (\log S + \gamma)
\end{equation}
(this is actually correct to quadratic order in $k$),
which when substituted in (\ref{bethe_eqn}) yields
\begin{eqnarray}
  k_n & = & \frac{2 \pi d}{N} I_n + \frac{2d}{N} (\gamma + \log S) 
        \sum_{m \neq n}(k_n - k_m) \nonumber \\
      & = & \frac{2 \pi d}{N} I_n + \frac{2d}{N} (\gamma + \log S) 
        (N k_n - k_{\rm tot}) \nonumber
\end{eqnarray}
and on substituting for $k_{\rm tot}$ from (\ref{ktot}) and rearranging,
we find
\begin{equation}
  \label{kn_solution}
   k_n = - \frac{\pi \left( I_n - \frac{\Sigma I_n}{N} \right)}{N(\gamma
            + \alpha)}.
\end{equation}

(Note that for very small $\eta$, $\alpha$ will be large and
negative, so the negative sign above is deceptive; the $k$'s are
ordered in the same way as the $I$'s.)
Equations (\ref{kn_solution}) thus give $k_n$ for any excited state
specified by any integers $I_n$, and the energy is $\sum k_n^2$ as
before. Note that the system now looks like a free Fermi gas or a
hard-sphere gas, which indeed is the underlying model behind the
asymptotic Bethe ansatz (we derived our results as a limiting case of
a gas of particles interacting by a $1/\sinh^2$ potential). There is
a continuous transition from this system to the classical Toda
lattice as $\eta$ is increased. As we show below, even in this limit
the excitations retain their qualitative features.

In the ground state, the $I_n$ are contiguous and may be taken to be
1, 2, \ldots, $N$. Then a simple calculation gives the ground state
energy as
\begin{equation}
  \label{e0_small_eta}
  E_0 = A N (N^2 - 1)/12 \approx AN^3/12
\end{equation}
where 
\begin{equation}
  \label{A}
  A = \frac{\pi^2}{N^2(\alpha + \gamma)^2}.
\end{equation} 

Now we consider excitations in which the last $l$ $I_n$ are excited
by an amount $m$---we insert $m$ holes between $I_{N-l}$ and
$I_{N-l+1}$, or in the phonon language, we add $m$ phonons in the
{$l$\/}th normal mode. $I_n$ are now 1, 2, 3, \ldots, $N-l$, $N-l+m+1$,
$N-l+m+2$, \ldots, $N+m+1$. Again, one can calculate the excitation
energy; it is
\begin{equation}
  \label{exc_small_eta}
  E - E_0 = A(Nl-l^2) \left( m + \frac{m^2}{N} \right).
\end{equation}

We consider several cases:

\begin{enumerate}
\item{
{\bf $m$ small, arbitrary $l$}
\nopagebreak

In this case, we get approximately
\begin{equation}
  \label{phononmode}
E - E_0 = A (Nl-l^2) m.
\end{equation}
This looks very much like a phonon dispersion; it rises from zero to
a maximum at the zone boundary, where its slope dies off. It 
is linear in the number of ``quanta'' $m$, and for the lower-energy 
modes (lower $l$) it is also linear in mode number or wavenumber
(i.e. the 2nd mode has twice
the energy of the first mode, and so on).

Moreover, for phonons we know that the zero point energy in each mode
is half the energy of one phonon; we can therefore sum half the above
expression over $l$, for $m=1$, and see, as a check, whether we
recover the zero point energy (\ref{e0_small_eta}). And indeed, we do
get
\[
\sum_{l-1}^{N-1} \half A(Nl - l^2) = \mbox{\small $\frac{1}{12}$} 
                   A N (N^2-1)
\]
in agreement with (\ref{e0_small_eta}).

The excitations are non-interacting---if we have several such
excitations in different modes their combined energy is the sum
of their individual energies, if there are not too many of them.
These hole excitations are thus quite analogous to phonons,
though they cannot be derived by approximating the lattice to a
harmonic lattice. 

}
\item{
{\bf $l=1$, $m$ large}
\nopagebreak

These are the excitations which one would expect to be soliton-like.
In this limit, we get
\begin{equation}
  \label{solitonmodes}
  E - E_0 = A(N-1) \left( m + \frac{m^2}{N} \right)
\end{equation}
For large $m$ the energy is thus quadratic in $m$. This energy,
however, is measured in the zero-momentum frame which is not the
frame in which one normally discusses solitons. The question of what is
the correct frame is discussed in the next section, where dispersion
relations are derived. 
}
\item{
{\bf $l$ small, $m$ large}
\nopagebreak

From (\ref{exc_small_eta}) we note that if $l$ is small, the
excitation energy is proportional to $l$. For instance, the energy
for $l=2$ is twice that for $l=1$. It is tempting to suppose that
this is a two soliton state, since the energies of solitons are
additive provided that they are few in number and hence well
separated ``most of the time''. In that case there would be a
continuous transition between a phononic excitation of the second
normal mode and the two soliton state, just as there is between the
excitation of the first normal mode and the one soliton state.
(Cf. Fig.\ \ref{cnoid}, and \S \ref{sec:classical})
}
\end{enumerate}
If the last two integers are excited by different amounts, one would
presumably have two solitons with different energies. Here, too, the
total excitation energy is the sum of the individual energies.
Carrying this picture further, an $N-1$ soliton state (with all
solitons having equal energies---$l=N-1$, $m$
large) has all the particles except one moving in one direction like
hard spheres, and is related by a Galilean transformation to a 1
soliton state. An $N$ soliton state (with all solitons
identical) is simply a uniform translation of the lattice as a whole. One
cannot put more than $N$ solitons in an $N$-particle lattice. The
last few sentences are speculative, but they indicate the possibility of
writing an arbitrary excited state as a kind of nonlinear
superposition of solitons. (To make this more convincing, read cnoidal
waves for solitons). Much the same thing is done
in the classical periodic system (\S \ref{sec:classical}).

\section{Dispersion relations for phonons and solitons}
\label{sec:dispersion}

We now find the dispersion relations for phonons and solitons.
First, however, we clarify the meaning of the momentum of
these excitations. 

As remarked earlier, the fact that we take the
dilute limit gives us a zero total momentum. Mertens' treatment,
on the other hand, yields a finite momentum $\sum k_n$
proportional to $\sum I_n$ and to the density $d$. This momentum is not 
a physically
relevant quantity. It is not the momentum of a phonon (though it is 
proportional to it), since it
depends on $\eta$ while the phonon momentum is a purely geometrical
quantity depending only on the system size and lattice spacing. Nor
is it the momentum of a soliton (it is not even proportional)
since the soliton momentum doesn't depend on the lattice spacing. 

The phonon momentum $q$ is the wavenumber of an
oscillatory excitation. For an $N$ particle
lattice $q$ has $N$ equally spaced values generally taken to lie
between $-\pi$ and $\pi$ (the first Brillouin zone)
in units of the inverse lattice spacing. The soliton
momentum is a little trickier to define in the quantum case. We
discuss it below. 

First consider the small $\eta$ limit.
We consider a single phonon, occupying normal mode $n$; its
excitation energy, from (\ref{phononmode}) with $l=n$ and $m=1$, is
$E-E_0 = A(Nn-n^2)$ and its wavenumber $q$, in units of inverse lattice
spacing, is $2\pi n/N$ (modulo $2\pi$;
we can choose the value to lie between $-\pi$ and $\pi$.) Note that
$\sum I_m = n$ in this case, if it was taken to be zero in the ground state, 
so $q$ is proportional to this quantity. This gives
$\omega$, the frequency (or the excitation energy of one phonon,
since $\hbar=1$) in terms of $q$ as
\begin{equation}
  \label{phonondisp_anharm}
  \omega = \frac{\pi}{2(\gamma +\alpha)^2} 
                   \left[ q - \frac{q^2}{2 \pi} \right],
              ~~~~0 < q < 2\pi
\end{equation}
and the phase velocity of sound is
\begin{equation}
  \label{phasevel_anharm}
   v_{\mbox{\scriptsize p}} = \frac{\pi}{2(\gamma + \alpha)^2}\left[ 1 -
                          \frac{q}{2\pi} \right]
\end{equation}
while the group velocity is
\begin{equation}
  \label{groupvel_anharm}
   v_{\mbox{\scriptsize g}} = \frac{\pi}{2(\gamma+\alpha)^2} \left[
                          1 - \frac{q}{\pi} \right].
\end{equation}
(in units of the lattice spacing).

\begin{figure}
  \begin{center}
 \leavevmode
    \epsfxsize=6.5cm 
\epsfbox{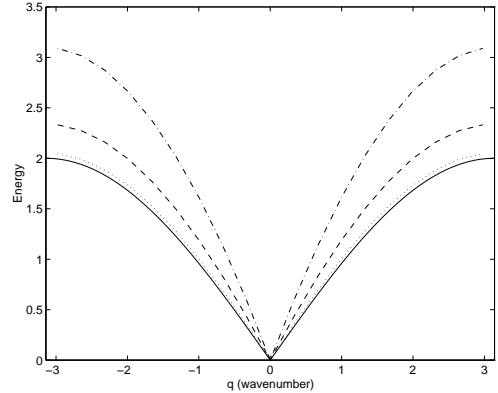}
 \end{center}
  \caption{Phonon energies (for the Hamiltonian (\ref{todaham_hbar},
            in units of $\hbar$), plotted against 
            wavenumber $q$ for various $\eta$. The solid line is the
            harmonic lattice curve and the curves for all ``large'' values of
            $\eta$ lie on top of it. The dotted line is for $\eta=2$, 
            the dashed line $\eta=0.1$, the dotdash line $\eta=0.01$.
            The range of $q$ is [$-\pi$,$\pi$].  Energies
           are in units of $\protect\sqrt{2\eta}$ (using the Hamiltonian
           (\ref{toda_ham}); or with the Hamiltonian (\ref{todaham_hbar})
           energies are in units of $\hbar$). 
            Here and in later graphs,
           units are chosen to get an $\eta$-independent curve in the
           large $\eta$ limit.
  \label{phonons}}
\end{figure}

\begin{figure}
  \begin{center}
  \leavevmode
  \epsfxsize=6.5cm
  \epsfbox{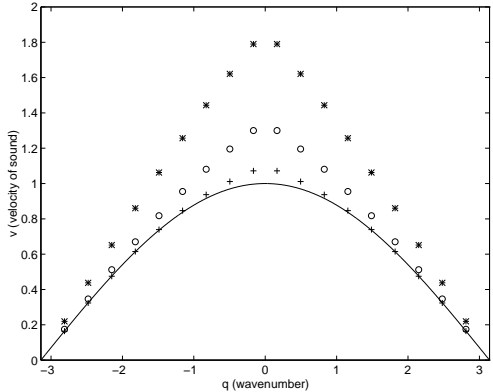}
  \end{center}
  \caption{The velocity of sound, $dE/dq$, plotted against $q$ for
           various $\eta$ for a 19 particle lattice. 
           The solid line is the curve for the harmonic 
           lattice, valid for large $\eta$. The crosses represent $\eta=2$,
           the circles $\eta=0.1$, the asterisks $\eta=0.01$. Units are
           as in the previous graph for the phonon dispersion.
   \label{soundvel}}
\end{figure}

\begin{figure}
  \begin{center}
  \leavevmode
  \epsfxsize=6.5cm
  \epsfbox{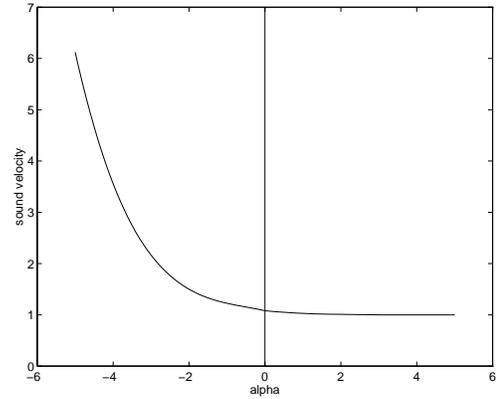}
  \end{center}
  \caption{Variation of long-wavelength sound velocity, in 
           the same units as in Figure \ref{soundvel}, as
           a function of $\alpha$ ($=-\log \hbar$ = $(1/2)\log (\eta/2)$).
           Note that $\alpha=0$ seems to divide the harmonic and
           quantum anharmonic regimes, i.e. the region where the
           harmonic approximation is valid for small excitations
           and the region where the zero point motion is so large
           that the harmonic approximation is not valid even in the
           ground state.
           \label{alphavel}}
\end{figure}

In the classical limit, of course, the phonons are what one would
find from a harmonic approximation. For a mode with wavenumber $q$
the energy is
\begin{equation}
E - E_0 = 2 \sqrt{2\eta} \sin \half q
\end{equation}
which yields the phase velocity (in units of lattice spacing)
\begin{equation}
  v_{\mbox{\scriptsize p}} = 2 \sqrt{2\eta} \frac{\sin \half q}{q}
\end{equation}
and the group velocity
\begin{equation}
 \label{groupvel_harm}
  v_{\mbox{\scriptsize g}} = \sqrt{2\eta} \cos \half q.
\end{equation}
The relations are different in the two cases, but have some similar
features, and at intermediate values of $\eta$ one obtains
interpolations between these. Dividing the
energies of excitation by $\sqrt{2 \eta}$ one gets results
independent of $\eta$ in the classical limit $\hbar \rightarrow 0$
or $\eta \rightarrow \infty$. 
 The results are plotted in Figs.\ 
\ref{phonons} and \ref{soundvel} (for a nineteen particle lattice).
One observes that for $\eta > 2$ the dispersion is more or less the
classical harmonic-lattice dispersion, while it begins to deviate for
$\eta < 2$. This is further emphasized by Fig.\ \ref{alphavel} which
shows how the long wavelength sound velocity varies with $\alpha =
- \log \eta$.

When we consider a soliton, we have to make clear what frame to view it
in to obtain an appropriate momentum. In the
classical case it is usually viewed in the frame where ``most'' of the
particles are at rest and only a localized excitation is moving. We
would like to choose a frame in the quantum case such that the dispersion
agrees with the classical formula; in particular as the energy of the
excitation increases it behaves more and more like a single hard sphere
moving in a stationary background and
the energy tends to $k^2$ (plus the ground state energy).

We identify a soliton with a state where $k_N$ is greatly excited
compared to all the other $k$'s. We can achieve the
$k^2$ dispersion if we work in a frame where the $k$'s excluding
$k_N$ are (roughly speaking) centred around zero.
In that case for large excitations $k_N \gg
k_n$ ($n<N$), the total momentum is very nearly $k_N$, the total
energy is nearly ${k_N}^2 \approx k^2$, and the quadratic dispersion
is achieved.

\begin{figure}
  \begin{center}
  \leavevmode
  \epsfxsize=6.5cm
  \epsfbox{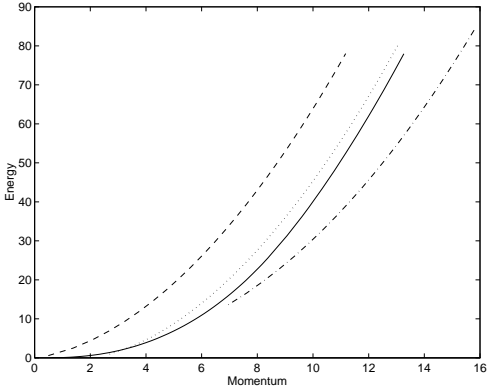}
  \end{center}
  \caption{Dispersion curves for the classical and quantum cnoidal waves.
           The solid curve is the classical cnoidal wave or soliton, viewed
           in the appropriate frame. The dotted curve is the quantum
           cnoidal wave in a frame in which $\sum_{n=1}^{N-1} k_n = 0$.
           This lies closest to the classical curve among the cases considered.
           The dashed curve corresponds to the frame in which only the ``inner''
           $k$'s are centred at zero, i.e. $\sum_{n=2}^{N-1} k_n = 0$ 
           (as in the ground state of the system).
           The dotdash curve corresponds to $k_N-1$ being fixed at its
           ground state value---implausible perhaps but included
           here for variety. All curves are for $N=10$ and
           (in the quantum case) for $\eta=1000$. The Hamiltonian 
           (\ref{todaham_hbar}) is used; in other words energies from the 
           Hamiltonian (\ref{toda_ham})
           are plotted in units of $\eta$.
   \label{solcquant}}
\end{figure}

However, exactly how to define the frame is not clear. There are
various possibilities---one could choose the average of all $k_n$
except $k_1$ and $k_N$ to be zero (so that the $k$'s are not very
much displaced from the ground state value); one could make the
average of $k_n$ including $k_1$ but excepting $k_N$ zero; one could
fix one of the $k$'s (say $k_1$, $k_{N/2}$ or $k_{N-1}$) to its 
ground state value; and so on. These possibilities are plotted in
Fig.\ \ref{solcquant}, for $\eta=1000$, and the 
dispersion for a classical cnoidal wave of
wavelength $N$ plotted for comparison, calculated
from the formula for $r_n = u_n-u_{n-1}$ given in \cite{todabook}
(cf.  \S \ref{sec:classical}). Of the possibilities listed
the second (where the $k$'s excepting $k_N$ average to zero) seems the
closest to the classical curve, but the agreement is imperfect and
the ``correct'' frame would appear to be something close but slightly
different. In plotting these curves we have used the 
Hamiltonian (\ref{todaham_hbar}),
whose limit as $\hbar \rightarrow 0$ is the classical problem in the correct
units. Fig.\ \ref{solcquant} shows the dispersion curves for $\eta=1000$.

\begin{figure}
  \begin{center}
  \leavevmode
  \epsfxsize=6.5cm
  \epsfbox{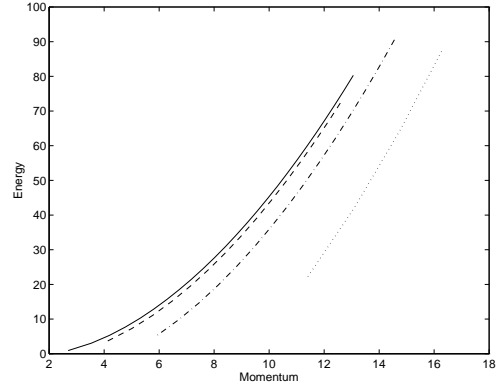}
  \end{center}
  \caption{The soliton dispersions, plotted in a frame in which 
           $\sum_{n=1}^{N-1} k_n = 0$. For $\eta>2$ all the curves
           lie on top of each other; they are shown by the solid line.
           The dashed line is for $\eta=2$ when they just start
           peeling apart. The dashdot line and the dotted line are
           for $\eta=0.1$ and $\eta=0.01$ respectively. The energies
           are in units of $\eta$, i.e. the Hamiltonian (\ref{todaham_hbar}),
           in terms of $\hbar$, is used.
   \label{solquant}}
\end{figure}

Fig.\ \ref{solquant} shows the particular
dispersion curve obtained by averaging $k_{n<N}$ to zero, for various
$\eta$. As in the case of the phonon curves, the soliton dispersions
lie on top of each other for large $\eta$ but begin peeling apart for
$\eta \approx 2$; as $\eta$ is reduced further they move further and
further away. Thus we find again that $\eta=2$ or $\hbar=1$
is a boundary between
classical and quantum regimes. For higher $\eta$ the dispersions are
essentially the classical ones apart from the discreteness of the
energy levels. For lower $\eta$ the results deviate significantly
from the classical ones. All the curves above have been calculated
for a ten particle lattice.

In the $\eta \rightarrow 0$ limit we
have the $k_n$ given by (\ref{kn_solution}); for the ground state
state we take the $I_n$ to be centred at zero (i.e. they range 
from $-(N-1)/2$ to $(N-1)/2$ for odd $N$, or from $-N/2$ to $N/2$ for
even $N$); and for the soliton we excite $I_N$ by an amount $m$.
Then $\sum I_n = m$. Clearly if we want the $k_n$ (for $n<N$) to be
centred at zero, we must add to (\ref{kn_solution}) a quantity
to cancel the $\sum I_n/N$ in the numerator, and instead subtract
$\sum_{n=1}^{N-1} I_n/(N-1)$. In this new frame, we
have 
\begin{eqnarray}
k_n & = & -\frac{\pi \left( I_n - \frac{\sum_{m=1}^{N-1} I_m}{N-1} \right)
                       }{N(\gamma+\alpha)}, \\
k   & = & \sum k_n = - \frac{\pi (m+N/2)}{N(\gamma + \alpha)}, \\
E-E_0 & = & \sum {k_n}^2 - E_0  \nonumber \\
  & = & \frac{\pi^2}{N^2(\gamma+\alpha)^2} \left[
                 m^2+Nm + \frac{N(N+1)}{4} \right].
\end{eqnarray}
The energy formula is not very different from (\ref{solitonmodes}).
The details of this formula should not be taken very seriously since
we are not clear about what the appropriate frame is in which to view the
soliton.  But the essential idea, that the energy is quadratic in the
momentum at large energies, will remain. In this frame the energy is
in fact (apart from a constant piece) purely quadratic in the
momentum---there is no linear term. This can be reconciled to our
picture of the low-$\eta$ limit as a hard sphere gas, so that at any
time the entire energy apart from the zero-point contribution comes
from the kinetic energy of one particle, the other particles being
at rest.

\begin{figure}
  \begin{center}
  \leavevmode
  \epsfxsize=6.5cm
  \epsfbox{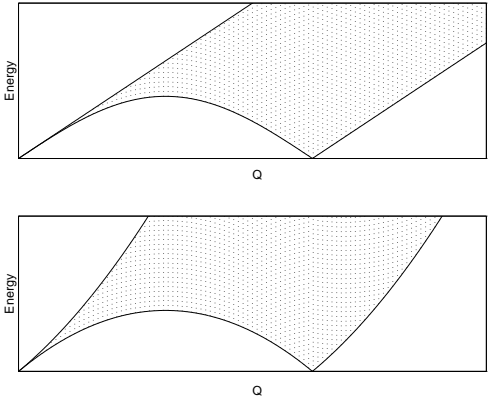}
  \end{center}
  \caption{Particle-hole excitations. In between the bounding upper and lower
           curves lies a continuum of allowed energy values corresponding
           to each $Q$ where $Q$ is as defined in \S \ref{sec:dispersion}
           for a single particle-hole pair. The upper graph corresponds to
           the harmonic limit, the lower graph to the $\eta \rightarrow
           0$ limit. (The energy scales are different in the two graphs.)
     \label{parthole}}
\end{figure}

Finally, if we wish to compare our system to the free Fermi gas
which it resembles in one limit, we could look at the ``particle-hole
excitation spectrum'' commonly plotted for such systems. To do this
we start from the ground state, with contiguous $I_n$; pick up one of
these, say $I_m$, move it to $I'_m$ (where $I'_m > I_N$ since all
other states are occupied), and define the momentum of this
``particle-hole excitation'' as $Q = 2 \pi (I'_m - I_m)/N$.
(This is basically the total phonon momentum of such an excitation.)
Then one gets a one-parameter range of energies for every $Q$, as
shown in Fig.\ \ref{parthole}. The 
harmonic and $\eta \rightarrow 0$ limits look
similar, qualitatively; the phonon or hole branch (the lower edge for
$Q<2\pi$) is a sine curve in the former case and a parabola in the
latter, and the particle branch (the upper edge and the lower edge
for $Q>2\pi$) is a straight line in the harmonic limit and a curve
(which indicates nonlinearity) otherwise.
The upper edge of the particle hole continuum has been identified with
a ``soliton'' by Sutherland, and corresponds to promoting
$k_N$ from the ground state configuration to one with a larger
value, and is thus essentially identical to our picture explained above.
A study of  the quantum numbers of the solitons
and the phonons
leads to a  suggestive  ``phonon decomposition''  of the soliton:
we can  view
the soliton  creation operator $A_q$ schematically  in  terms of a phonon creation operator
$a^\dagger_q$ as
\begin{equation}
A^\dagger_{\frac{2 \pi}{N} m} \; \; \sim \;\; [a^\dagger_{\frac{2 \pi}{N}} ]^m ,
\end{equation}
i.e. a particular kind of highly symmetric multi phonon state.

\section{ Correlation Functions, Finite-size Effects and Conformal Theory}
\label{sec:EvsN}

We now turn to the issue of correlation functions of the
Toda lattice,  making  contact with the the theory of conformal
invariance in this class of systems. Conformal invariance
has given considerable insight into correlation functions
of quantum many body  models having critical behaviour, as typified by a vanishing
of excitation energies or power law correlations, and useful reviews
of this fast growing field are to be found in \cite{christe}
and in \cite{kawakami}.

Let us first note that the quantum Toda lattice in its ground  state
is not quite a lattice: the Bragg peaks are melted due to 
zero point motion. In the harmonic limit this is simple to
see, since we can write the displacement in terms of the 
phonon creation operators and the phonon dispersion
$\omega_q = 2 \; v \; |\sin(q/2) |$ as
\begin{equation}
u_n=\frac{1}{\sqrt{N}} \sum_q \; \exp(i q n) \; \;
 \frac{1}{i \sqrt{\omega_q}} \; \; (a^\dagger_q - a_{-q}) 
\end{equation}
whereby $\left< u^2_n \right> = (1/N)  \sum (1/\omega_q) \; \sim 
(1/\pi v) \log(N)$. The  phonon velocity $v = \sqrt{2 \eta}$ in the
harmonic limit of the Toda problem. The structure function at the
first reciprocal lattice vector $G = (2 \pi /N)$ is 
\begin{eqnarray}
\left<\rho_G \; \; \rho_G \right> & = & \sum_{m,n} \left< 
    e^{ i 2 \pi u_n} \; \; e^{ -i 2 \pi u_m} \right> \nonumber \\ 
\left< e^{ i 2 \pi u_n} \; \;  e^{ -i 2 \pi u_m}\right> & =&  
    e^{- 2 \pi^2 \langle (u_n-u_m)^2 \rangle } 
          =  e^{- 4 \pi /v \log(|m-n|)  } \nonumber \\
          &=& \frac{1}{|m-n|^{4/\pi v} } \label{correlation}
\end{eqnarray}
where we have used the Gaussian cumulant theorem 
$\left< \exp(a) \right> =\exp(1/2 \left< a^2 \right> )$
and the logarithmic integral $(1/N) \sum(1- \cos(q r))/\omega_q \sim
(1/\pi v) \log(r/r_0)$.  We thus see that the Toda lattice may be
expected to have powerlaw correlations for all $\eta$,  since
it has low energy excitations for all $\eta$, namely the phonons. 

A  characteristic of conformally invariant theories is the 
`central charge' $c$. One way of  checking for conformal invariance
is to compute corrections to the ground state energy for a finite
sized system, which is expected to have a behaviour
\begin{equation}
E(L) = L e_\infty  - \frac{c  \pi v}{6 L} + O(1/L^2)  \label{energycentral}
\end{equation}
where $v$ is the velocity of the low lying excitations, such that
a tower of excited states exists with energy $v 2\pi/L \times \mbox{integer}$.
A glance at (\ref{harmoniclargen}) shows that in that limit of large
$\eta$ we have $c=1$, as indeed does the initial $1/\sinh^2$ model. The case
of $c=1$ usually leads to exponents varying continuously with coupling
constants, and hence (\ref{correlation}) is consistent with this possibility.
In the present model, we must,  however, first establish that
the asymptotic Bethe ansatz gives the correct  energy to O($1/N$) or
O($1/L$). This is not guaranteed {\em a priori\/} by any theoretical argument
and must be checked for self consistency. (Incidentally, in the
Toda lattice we are at a fixed density so we will not distinguish
between $L$ and $N$.) The internal check performed is to compute
the velocity at a fixed $\eta$ and to compute the energy for 
various $N$ and to check against Eq.\ (\ref{energycentral}).

First we note that in the extreme anharmonic limit
equation (\ref{e0_small_eta}) for the ground state in the
low $\eta$ limit does indeed give the same sound
velocity as (\ref{phasevel_anharm}) or (\ref{groupvel_anharm}), so in the 
low $\eta$ limit $c=1$  exactly, as it is in the harmonic limit.

We performed the calculation  for $\eta=2, 10, 100$ 
 (Table \ref{tab:EvsN}). As in figures \ref{phonons} and
\ref{soundvel}, we use the Hamiltonian (\ref{todaham_hbar})
and units of $\hbar$ (equivalently, the Hamiltonian
(\ref{toda_ham}) with units of $\sqrt{2\eta}$); in these units the
sound velocity for the harmonic lattice is 1 exactly.
From these results, we get
\begin{tabbing}
$ vc  =$ \= $1.0764 \pm 0.0006$ \= ($\eta=2$) \\
                \> $1.035 \pm 0.003$   \> ($\eta=10$) \\
                \> $1.01 \pm 0.03$     \> ($\eta=100$) 
\end{tabbing}
On interpolating the 19-particle results of Fig.\ \ref{soundvel} for q=0,
we get the estimates $v=1.08, 1.04, 1.01$ for $\eta=2, 10, 100$
respectively, with uncertainties in the second decimal place.
Thus we get for the central charge 
\begin{tabbing}
$c =$ \= $1.00 \pm 0.01$ \= ($\eta=2, 10$) \\
      \> $1.00 \pm 0.03$  \> ($\eta=100$).
\end{tabbing}
The uncertainty in the cases $\eta=2$, $10$ arises mainly from the
inaccuracy in our determination of $v$.
The results seem to indicate that 
$c$ is equal to 1 at all values, and moreover it is
reproduced correctly by the Bethe ansatz even at $\eta=100$, which is
well into the ``classical'' limit.
 It thus appears that  the error in
energy per particle goes, at worst, as the inverse cube of the number
of particles. The error bars could be reduced by increasing the
system size further.

In the anharmonic limit, in fact, the series stops there ($E/N$ has only
a constant piece and a $1/N^2$ piece) while in the harmonic limit all
odd powers, $1/N^3$, $1/N^5$ and so on are missing. One might conjecture that
this is the case at all values of $\eta$.
For $\eta=2$ we took the ground state energies per particle for various $N$,
subtracted $e_\infty$ and the $1/N^2$ piece, and fitted the results
to power series in $1/N$ starting at $N^{-3}$. The result was
a coefficient of $0.021 \pm 0.006$ for the $N^{-3}$ term, and
$-1.1 \pm 0.2$ for the $N^{-4}$ term. Thus the coefficient of the $1/N^3$ term
does seem to be very nearly zero. For $\eta=10$ and $100$ the numbers
we obtained didn't allow us to make such fits---the error bars turned
out to be much larger than the values themselves. We conjecture
that the coefficient of the $1/N^3$ term vanishes at all
$\eta$, but for high $\eta$ the Bethe ansatz may not be accurate to
this order in $N$ and may be unable to reproduce this result. We
are unable to make a statement about higher odd powers.

\begin{table}
 \caption{Ground state energy per particle as a function of system size
   \label{tab:EvsN}}
 \begin{tabular}{|c|ccc|} \hline
N & & $E/N$ & \\
  & $\eta=2$ & $\eta=10$ & $\eta=100$ \\ \hline
29  &  1.675512397777  &  2.890224040772  &   \\ 
33  &  1.675665073759  &  2.890370838890  &   \\ 
41  &  1.675847391894  &  2.890546128059  &   \\ 
49  &  1.675947956073  &  2.890642813964  &   \\ 
57  &  1.676009234239  &  2.890701728785  &  7.713369630480  \\ 
65  &  1.676049312911  &  2.890740261674  &  7.713407380691  \\ 
81  &     &   &   7.713452036110  \\ 
97  &     &   &   7.713476466168  \\ 
113  &     &   &   7.713491273955  \\ 
129  &     &   &   7.713500922044 
\end{tabular}
\end{table}

Accepting that the Toda lattice is a $c=1$ theory, we can 
establish the power law of the density correlator as in (\ref{correlation}),
without too much detailed calculation, on using the Galilean
invariance of the model. The theory of conformal invariance 
(see e.g. \cite{kawakami})
says that if we have an excitation that boosts the total momentum
by $k_{tot}$ then the change in energy is
\begin{eqnarray}
\delta E &  = &  2 \pi v x/N  \\
x & = & (\frac{k_{tot}}{2 \pi})^2 \mu \\
\alpha & =& 2 \mu
\end{eqnarray}
where $\alpha$ is the exponent determining the decay of a 
primary operator.  However, Galilean invariance implies that
\begin{equation}
\delta E = \frac{k_{tot}^2}{N};
\end{equation}
hence we find
\begin{equation}
\alpha = \frac{4 \pi}{v}.
\end{equation}
Comparing with the harmonic limit result (\ref{correlation}),
we see that the primary operator may be identified with
the density fluctuation $\rho_G$ and hence the result 
(\ref{correlation}) is true at all $\eta$ provided we
substitute the appropriate value of $v(\eta)$.  A similar result
is well known to be true for the $1/r^2$ models for the density
correlation function, but unlike in that case, there 
is a difficulty in defining a ``bosonic'' correlator,
since we are always working at a fixed density, and hence
the compressibility is zero.

\section{Comparison with the classical Kac - Van Moerbeke formulation}
\label{sec:classical}

To summarize the above, we now have a picture of how the $k$'s in the Bethe
ansatz (or the $\rho$'s in Gutzwiller's treatment) behave, in the
ground state and in the excited states. In the ground state the $I$'s
and therefore the $k$'s
are all closely spaced. In the excited states the separations between
them widen. If there is a gap of $m$ integers
between $I_{N-n}$ and $I_{N-n+1}$ the
gap between $k_{N-n}$ and $k_{N-n+1}$ widens and
one has $m$ phonons in the $n$th normal mode. If the gap between the
$k$'s becomes very
large the excitation becomes solitonic. In
particular, for $n=1$ one has a one-soliton state; for $n=2$, a two
soliton state (with equal amplitudes); and so on.

We now compare this description with the description
of the system in the classical variables of Kac and van Moerbeke
\cite{KvM,todabook}. Briefly they use the
variables $\mu_1, \mu_2, \ldots, \mu_{N-1}$ which are the eigenvalues
of a truncated Lax matrix obtained by striking off the first row and
column (i.e. removing the first particle from the problem). These
$\mu$'s are the momenta of the particles in the remaining open chain 
if the system is dilute. Kac and van Moerbeke show that these $\mu$'s
are confined to the $N-1$ closed intervals where the characteristic
polynomial of the Lax matrix, $\left| \lambda I - L \right| $
is equal to or greater than 2 in magnitude. The polynomial goes to
$\pm \infty$ for large $\lambda$, while it oscillates in the middle;
for the ground state it touches the lines $\lambda = \pm 2$ in $N-1$
places so that the closed intervals referred to above are single
points and all the $\mu$'s are stationary. For an excited state the
polynomial crosses the lines $\lambda = \pm 2$ so the closed
intervals get a finite width and the $\mu$'s oscillate inside these
intervals as the system evolves. 

The analogues of the classical $\mu$'s are Gutzwiller's $\rho$'s, or
approximately Sutherland's $k$'s. Whereas there are $N-1$ $\mu$'s
each confined to a different interval in the classical picture, in
Gutzwiller's picture each of the $N-1$ analogous variables has a
spectrum of $N$ values $\rho_n$. On calculating the classical $\mu$'s
in the ground state, as is done in \cite{todabook}, we find that their 
values lie almost
exactly in between the quantum $\rho$ (ie $k$) values. There is an analogy
between the $\mu$'s and the ``gaps'' in the $k$-spectrum. In the ground state
the gaps are minimum, the $\mu$'s fit into these gaps, and the $\mu$'s are
stationary. In an excited state some or all of these gaps between the $k$'s
widen, and the corresponding $\mu$'s are no longer stationary but oscillate in
intervals of finite width. In particular a pure cnoidal wave corresponds to
exactly one $\mu$ acquiring a width in which to oscillate, 
or exactly one gap among the $I_n$ (hence the $k_n$) widening.

A single cnoidal wave has the formula \cite{todabook}
\begin{equation}
  e^{-r_n} = 1 + (2 K \nu )^2 \left( \mbox{dn}^2 \left[
                    2(n/\lambda \pm \nu t)K \right] -E/K \right),
\end{equation}
where $r_n = u_n-u_{n-1}$, $K$ and $E$ are the complete elliptic
integrals of the first and the second kind, $\lambda$ is the
wavelength ($=N$ for the first ``normal mode'' or one soliton, $N/2$
for the second normal mode, etc) and $\nu$ is given by
\begin{equation}
2 K \nu = \left[ \frac{1}{\mbox{sn}^2(2K/\lambda)} - 1 + \frac{E}{K} 
              \right] ^{-1/2}
\end{equation}

For low modulus $k$ of the elliptic functions, this is like a
sinusoidal wave, but as the modulus increases it becomes sharply
peaked locally and flat elsewhere (Fig.\ \ref{cnoid}). 
As remarked in \S \ref{sec:dispersion},
the dispersion calculated from this expression is close to the
dispersion, in an appropriate reference frame, of the quantum cnoidal
wave.

\section{Conclusions}

In conclusion, we have shown that the usefulness of the asymptotic
Bethe ansatz in the quantum Toda problem is not confined to finding
thermodynamic properties. The method gives results for energy
per particle accurate to
O($1/N^2$), which is sufficient to calculate finite size effects and even
correlation functions using conformal theory. The O($1/N^3$) term
seems to vanish in the exact solution, though the Bethe ansatz solution
probably doesn't reproduce this result.

We have demonstrated that in fact the Bethe ansatz equations are a 
simplification of Gutzwiller's
method and can be derived from them. The parameter governing the
error can be taken to be the difference in $\rho_n$ and $\epsilon_n$
in \S \ref{sec:Gutzwiller}. According to Matsuyama \cite{matsu96} this
difference falls exponentially with $N$, so that the error
goes as $e^{-N/f(\eta)}$, where $f(\eta$) is some dimensionless number.
We also show that the error vanishes as $\eta$ becomes small, so that
$f(\eta) \rightarrow 0$ as $\eta \rightarrow 0$.

Thus, we can treat finite sized systems,
account for low-lying states (phonons) and higher excitations
(solitons), and find their dispersions and velocities. Comparison
with conformal theory gives the ``central charge'' $c=1$, which means 
that the coefficient of the $1/N^2$ term in the $E/N$ expansion is
essentially the sound velocity.

We find that the properties of excitations 
are very similar to the classical 
properties for $\eta>2$
($\hbar < 1$), apart from the underlying discreteness of the energy
levels. The quantization is then analogous to the quantization of a
harmonic lattice. The soliton, which is an effect of large occupation of one
mode, is no different from the classical object described by Toda; even its
energy is effectively not quantized since the occupation number is so large.

For small $\eta$ (large $\hbar$) things are different: the phonons
no longer derive from a harmonic approximation, and the soliton dispersions
no longer match the classical ones,  though
qualitatively the dispersion curves retain some similar features,
both for solitons (high-amplitude cnoidal waves) and for phonons.
For both excitations the dispersions depend on $\eta$, and moreover the energy
of a mode deviates rapidly from linearity with increasing occupation number
$n$, so that $n$ need not be macroscopic (at least for finite lattice size $N$)
for the mode to become soliton-like---the soliton's energy is indeed
quantized. Thus if the large $\eta$ soliton is
essentially the soliton of Toda's classical lattice, the corresponding
small $\eta$ object deserves to be called the quantum soliton.

\appendix

\section*{H\'{e}non's integrals, classical and quantum}

In this appendix we discuss the integrability of the Toda lattice
classically and quantum mechanically; while much of the discussion is
not new it seems difficult to find it in one place elsewhere.
Following Pasquier and Gaudin \cite{pasq-gaud}, who give a proof of
quantum integrability, we show that their conserved quantities are the same
as H\'{e}non's integrals---whose conservation is necessary for
Gutzwiller's treatment to go through.

The equations of motion for the classical lattice can be written
in the Lax form
\begin{equation}
\frac{dL}{dt}=LM-ML, \label{laxeqn}
\end{equation}
where
\begin{eqnarray}
L= \left( \begin{array}{ccccc} 
         b_1 & a_1 & & & a_N \\
         a_1 & b_2 & a_2  \\
         & a_2 & b_3 \\
         & & & \ddots & a_{N-1} \\
         a_N & & & a_{N-1} & b_N
         \end{array} \right), \\ [0.5em]
M= \left( \begin{array}{ccccc} 
         0 & a_1 & & & -a_N \\
         -a_1 & 0 & a_2  \\
         & -a_2 & 0 \\
         & & & \ddots & a_{N-1} \\
         a_N & & & -a_{N-1} & 0
         \end{array} \right) 
\end{eqnarray}
and
\begin{equation}
a_j = e^{-(q_{j_1}-q_j)/2}, ~~~b_j=p_j.
\end{equation}

From this one can show that the eigenvalues of the Lax matrix $L$, or
equivalently the coefficients $I_n$ of the characteristic
polynomial of the Lax matrix are conserved quantities \cite{todabook}. 
These are H\'{e}non's integrals, and are given by
\begin{equation}
I_m = \sum_{{i_1, i_2, \ldots, i_k,} \atop {j_1, j_2,\ldots, j_l}}
  p_{i_1} p_{i_2} \ldots p_{i_k} \left(-X_{j_1} \right)
    \left(-X_{j_2} \right) \ldots \left(-X_{j_l} \right)
\end{equation}
where
\begin{equation}
 X_j = e^{-(q_{j+1}-q_j)},
\end{equation}
there are no repeated indices in the $p$'s or the $q$'s in a given
term ($i_1$, $i_2$, \ldots, $j_1$, $j_1+1$, $j_2$, $j_2+1$, \ldots are all
different), the total number of such indices in each term is $m$ 
(i.e. $k+2l = m$) and the sum is over all distinct terms satisfying these
conditions (i.e. terms not differing merely in the order of factors).

In quantum mechanics, the coefficients of the Lax matrix are the
same, and have no ordering problems, but now the equations of
motion (\ref{laxeqn}) are no longer valid (each term in the matrix product
has to be ordered)
and the proof that the coefficients are conserved fails.
Gutzwiller \cite{gutz} assumes that they are conserved
nonetheless (he only takes the cases $N=3,4$ where it can be
verified easily). Their conservation can be shown as a 
consequence of 
the work of Pasquier and Gaudin \cite{pasq-gaud}, who prove that
the coefficients of $u$ in the trace of the ``monodromy matrix'' $T_N$
are in involution, where
\begin{eqnarray}
T_N(u) & = & L_1 L_2 \ldots L_N, \\
L_n(u) & = & \left( \begin{array}{cc}
                     u-p_n & \; e^{q_n} \\[0.5em]
                     - e^{-q_n} & \; 0 
                    \end{array} \right).
\end{eqnarray} 
The definitions hold in both the classical and the quantum
cases. Classically their conservation follows from the classical
equations of motion
\begin{eqnarray}
\frac{dL_n}{dt} & = & M_{n-1}L_n - L_n M_n, \\
\mbox{where} 
  ~~M_n & = & \left( \begin{array}{cc}
                     u & \; e^{q_n} \\[0.5em]
                     - e^{-q_{n+1}} & \; 0 
                    \end{array} \right).
\end{eqnarray}
Quantum mechanically these satisfy the Yang-Baxter equations:
we may rewrite $L_n \rightarrow L_{n,g}(u)= (u- p_n) (1+\sigma_g^z)/2
- \exp(-q_n) \; \sigma_g^- + \exp(q_n) \;\sigma_g^+ $ and show that the 
monodromy matrix $T_N(u) \rightarrow T_{g}(u)$ satisfies the Yang-Baxter 
condition 
$T_g(u) T_{g'}(v) R_{g,g'}(u-v) =R_{g,g'}(u-v) T_{g'}(u) T_g(v)$
with  $R_{g,g'}=a(u-v) + b(u-v) \vec{\sigma_g}. \vec{\sigma_{g'}}$.
Taking a trace  over the auxiliary spaces $\sigma_g, \sigma_{g'}$
the integrability is established. We now show that  
these coefficients are in fact H\'{e}non's integrals.
Consider a polynomial in $u$, $F_N(u)$, defined by
\begin{eqnarray}
F_N(u)  = \sum^{k+2l=N} & &(u-p_{i_1}) (u-p_{i_2}) \ldots (u-p_{i_k})
   \nonumber \\
    & &\ldots (-X_{j_1})(-X_{j_2}) \ldots (-X_{j_l})
\end{eqnarray}
where the indices satisfy the same restrictions as in the
definition of H\'{e}non's integrals. It is easily seen that
\begin{equation}
F_N(u) = \sum_{n=0}^N (-1)^n I_n u^{N-n}.
\end{equation}
We can show by induction that this polynomial is the trace of 
$T_N(u)$. Defining
\begin{eqnarray}
F'_N(u) & = & \mbox{All the terms in $F_N(u)$} \nonumber \\
   & & \mbox{ which do not include
           a factor $e^{q_N}$}, \\
F''_N(u) & = & F_N(u) - F'_N(u) \nonumber \\
      & = & \mbox{All the terms in $F_N(u)$} \nonumber \\
      & & \mbox{ which include a factor $e^{q_N}$},
\end{eqnarray}
we claim that
\begin{equation}
T_N = \left( \begin{array}{cc}
                     F'_N(u) &\; e^{q_N} F'_{N-1}(u) \\[1em]
                     e^{-q_{N+1}}F''_{N+1}(u) &\; F''_N(u)
                    \end{array} \right).
\end{equation}
The claim is easily verified for $N=1,2,$ etc. Suppose it is
true for $N$; then
\begin{eqnarray}
&& T_{N+1}  =  T_N L_{N+1} = \nonumber \\ [0.5em]
&& ~~  \left( \begin{array}{cc}
 (u-p_{n+1})F'_N(u) \; \;   & e^{q_{N+1}} F'_N(u) \\
  - e^{q_N-q_{N+1}}F'_{N-1}(u) & \\[1.5em]
   (u-p_{N+1})e^{-q_{N+1}} F''_{N+1}(u) \;
                      \; & F''_{N+1}(u) \\
  -e^{-q_{N+1}} F''_N(u) & 
                    \end{array} \right)
\end{eqnarray}
which one can check is the same as
\begin{equation}
T_{N+1} = \left( \begin{array}{cc}
              F'_{N+1}(u) & e^{q_{N+1}} F'_N(u) \\[1em]
              e^{-q_{N+2}}F''_{N+2}(u) & F''_{N+1}(u)
                    \end{array} \right).
\end{equation}
Thus our claim is true for all $N$, and in particular the trace of $T_N$ is
$F_N(u)$.

\end{document}